\documentclass[aps,prd,twocolumn,superscriptaddress,preprintnumbers,floatfix,nofootinbib,notitlepage,showkeys,showpacs]{revtex4-1}

\usepackage[utf8]{inputenc}
\usepackage{cancel}

\usepackage{graphicx}
\usepackage{hyperref}
\usepackage{latexsym}
\usepackage{amsmath}
\usepackage{amssymb}
\usepackage{bbm}

\usepackage[normalem]{ulem}
\usepackage{pdfsync}
\usepackage{epsfig}
\usepackage{epstopdf}
\usepackage{subfigure}
\usepackage{color}
\usepackage{comment}
\usepackage{slashed}
\usepackage{placeins}
\usepackage{cleveref}

\newcommand{\be}{\begin{equation}} 
\newcommand{\ee}{\end{equation}}
\newcommand{\bea}{\begin{eqnarray}} 
\newcommand{\eea}{\end{eqnarray}}

\newcommand{\bmp}{\noindent\begin{minipage}{16cm}}
\newcommand{\emp}{\end{minipage}\vskip 7mm} 
\def\lsim{\mathrel{\raise.3ex\hbox{$<$\kern-.75em\lower1ex\hbox{$\sim$}}}}
\def\gsim{\mathrel{\raise.3ex\hbox{$>$\kern-.75em\lower1ex\hbox{$\sim$}}}}

\newcommand{\intron}[1]{}

\usepackage{xcolor}
\definecolor{cites}{RGB}{0,180,0}
\definecolor{links}{RGB}{200,0,0}
\hypersetup{colorlinks=true,citecolor=cites,linkcolor=links,urlcolor=blue}

\newcommand{\kr}{}

\begin{document}

\title{From chiral symmetry breaking to conformality in SU(2) gauge theory}

\author{Alessandro Amato}
\email{alessandro.amato@helsinki.fi}
\affiliation{Department of Physics, University of Helsinki \\
                      P.O.~Box 64, FI-00014, Helsinki, Finland}
\affiliation{Helsinki Institute of Physics, \\
                      P.O.~Box 64, FI-00014, Helsinki, Finland}

\author{Viljami Leino}
\email{viljami.leino@helsinki.fi}
\affiliation{Department of Physics, University of Helsinki \\
             P.O.~Box 64, FI-00014, Helsinki, Finland}
\affiliation{Helsinki Institute of Physics, \\
             P.O.~Box 64, FI-00014, Helsinki, Finland}

\author{Kari Rummukainen}
\email{kari.rummukainen@helsinki.fi}
\affiliation{Department of Physics, University of Helsinki \\
                      P.O.~Box 64, FI-00014, Helsinki, Finland}
\affiliation{Helsinki Institute of Physics, \\
                      P.O.~Box 64, FI-00014, Helsinki, Finland}

\author{Kimmo Tuominen}
\email{kimmo.i.tuominen@helsinki.fi}
\affiliation{Department of Physics, University of Helsinki \\
                      P.O.~Box 64, FI-00014, Helsinki, Finland}
\affiliation{Helsinki Institute of Physics, \\
                      P.O.~Box 64, FI-00014, Helsinki, Finland}

\author{Sara T\"ahtinen}
\email{sara.tahtinen@helsinki.fi}
\affiliation{Department of Physics, University of Helsinki \\
                      P.O.~Box 64, FI-00014, Helsinki, Finland}
\affiliation{Helsinki Institute of Physics, \\
                      P.O.~Box 64, FI-00014, Helsinki, Finland}


\begin{abstract}
We compute the spectrum of SU(2) gauge theory with two, four or six Dirac
fermions in the fundamental representation of the gauge group.
We investigate the scaling of the meson masses as a function of the quark
mass when approaching the chiral limit. We find behavior compatible with the
usual pattern of chiral symmetry breaking at $N_f\le 4$ and with the existence
of an infrared fixed point at $N_f=6$. In the six flavor case we determine the
anomalous dimension of the quark mass operator from the spectrum and find
results fully consistent with earlier analyses.
\end{abstract}

\preprint{HIP-2018-12/TH}

%
\maketitle
%
\section{Introduction}
\label{sec:intro}
The Standard Model (SM) is incomplete, as it cannot explain the origin
of dark matter or the observed matter-antimatter asymmetry. The fine-tuning
and naturalness problems of the scalar sector of the SM provide
further motivations for constructing beyond SM theories.
These issues can be addressed within models where
a new strong dynamics underlies the electroweak symmetry
breaking of SM~\cite{Susskind:1978ms,Weinberg:1979bn,Hill:2002ap,Sannino:2008ha}.

Constructing this type of theories provides challenges for our
understanding of strong dynamics in general. A particularly
important problem is the determination of the vacuum phase of
gauge theory SU($N$) and including some number $N_f$ of fermions
in some representation of the gauge group\cite{Sannino:2004qp}

Consider SU($N$) with fermions in the fundamental representation.
As a function of $N_f$, there are several distinct features:
Asymptotic freedom is lost above $N_f=11 N/2$, and
directly below this value exists the conformal window. Within the conformal window
the long distance behavior of the theory is governed by an infrared fixed point (IRFP).

The existence of a weakly coupled IRFP can be rigorously established in
the limit of large number of colors and flavors~\cite{Banks:1981nn}.
Nonperturbatively, and for finite $N$, the conformal window is expected to extend to some
critical value $N_{f}^{\rm{crit}}$. If $N_f$ is further decreased, the long distance vacuum
properties of the theory become akin to QCD: at zero quark
masses the chiral symmetry of fermions breaks spontaneously
leading to the appearance of a corresponding
multiplet of Goldstone bosons in the physical spectrum.

A simple template model framework where all these dynamics can be realized
is provided by SU(2) gauge theory with $N_f$ massless Dirac fermions in the
fundamental representation of the gauge group~\cite{Sannino:2004qp}.
These theories have been studied on the lattice by several
groups~\cite{Ohki:2010sr,Bursa:2010xn,Karavirta:2011zg,Lewis:2011zb,Hayakawa:2013yfa,Appelquist:2013pqa,Hietanen:2014xca,
Leino:2017lpc,Leino:2017hgm}. There is evidence that
the theory with 10 and 8 flavors has a fixed point~\cite{Karavirta:2011zg,
Leino:2017lpc}
and the theory with
$N_f=2$ and 4 is outside the conformal window and breaks chiral symmetry
according to the expected pattern. The case $N_f=6$ has long remained inconclusive~\cite{Bursa:2010xn,Karavirta:2011zg,Hayakawa:2013yfa,Appelquist:2013pqa}.
but recent results indicate that it is within the conformal window \cite{Leino:2017hgm,Leino:2018qvq}.

In this paper we report the results of a systematic study of the spectrum of
the SU(2) gauge theory with $N_f=2$, 4 and 6 flavors.
We measure the hadron masses as a function of the quark mass, and the two and four flavor cases are observed to follow the usual pattern of chiral symmetry breaking when approaching the chiral limit. We find that the six-flavor
theory does not follow similar behavior. The scaling of the masses with
respect to the quark mass is
compatible with the interpretation that the six flavor theory is inside but
near the lower boundary of the conformal window.

From the dependence of the hadron masses on the quark mass, we also extract the
value of the anomalous dimension of the quark mass operator in the six flavor
case. We compare our results on the mass anomalous dimension with the recent
analysis in \cite{Leino:2017hgm} where the step-scaling method was applied, and
find that the two are fully compatible with each other, \kr{within the statistical errors,} lending further support
for the conclusion that the SU(2) gauge theory with six flavors is inside the
conformal window.

The paper is organized as follows: We introduce the lattice model we study in section \ref{sec:model} and show the results of our numerical analysis in
section~\ref{sec:results}. In section~\ref{sec:chekout} we present our conclusion.

\section{Lattice Model}
\label{sec:model}

We investigate SU(2) gauge theory with $N_f$ Dirac fermions in the fundamental representation of the gauge group on the lattice. The lattice action is
\be
S=(1-c_g)S_G(U)+c_gS_G(V)+S_F(V),
\label{eq:action}
\ee
where $S_F$ is the HEX-smeared~\cite{Capitani:2006ni} clover improved Wilson fermion action and
$S_G$ is the plaquette gauge action. The smeared and unsmeared gauge fields are denoted by $V$ and
$U$, respectively. The gauge action
smearing is set to $c_g=0.5$. The partially smeared gauge action allows one to avoid the unphysical bulk phase transition
at strong coupling~\cite{DeGrand:2011vp} and the details of the smearing we use are described in~\cite{Rantaharju:2015yva}.  \kr{Similar action was used
in~\cite{Leino:2017hgm}, where the evolution of the coupling constant in SU(2) gauge theory with $N_f=6$ fermions was measured.}

In Eq.~\eqref{eq:action} $S_F(V)$ is the usual clover improved Wilson fermion action and the
Sheikholeslami-Wohlert coefficient is set equal to one.
The approach to the chiral limit is investigated by tuning the physical quark mass, defined by the lattice PCAC relation~\cite{Luscher:1996vw}
\be
aM(x_0)=\frac{(\partial_0^\ast+\partial_0)f_A(x_0)}{4f_P(x_0)},
\ee
to zero. Here the axial current and density correlation functions are
\begin{align}
f_A(x_0) &=-\frac{1}{12L^6}\int d^3yd^3z\langle\bar\psi(x_0)\Lambda_{05}^a\psi(x_0)\bar\zeta(y)\Lambda_{05}^a\zeta(z)\rangle,\notag \\
f_P(x_0) &=-\frac{1}{12L^6}\int d^3yd^3z\langle\bar\psi(x_0)\Lambda_5^a\psi(x_0)\bar\zeta(y)\Lambda_5^a\zeta(z)\rangle\,\notag
\end{align}
where $\Lambda_{05}^a=\gamma_0\gamma_5\lambda^a$ and $\Lambda_5^a=\gamma_5\lambda^a$ in terms of Dirac gamma matrices and SU($N_f$) generators $\lambda^a$, \kr{which act on the flavor components of the spinors.}
The masses of color singlet meson states are determined by fitting the time sliced average correlation functions with Coulomb gauge fixed wall sources.

\section{Measurements and results}
\label{sec:results}
\begin{figure}
\includegraphics[width=.45\textwidth]{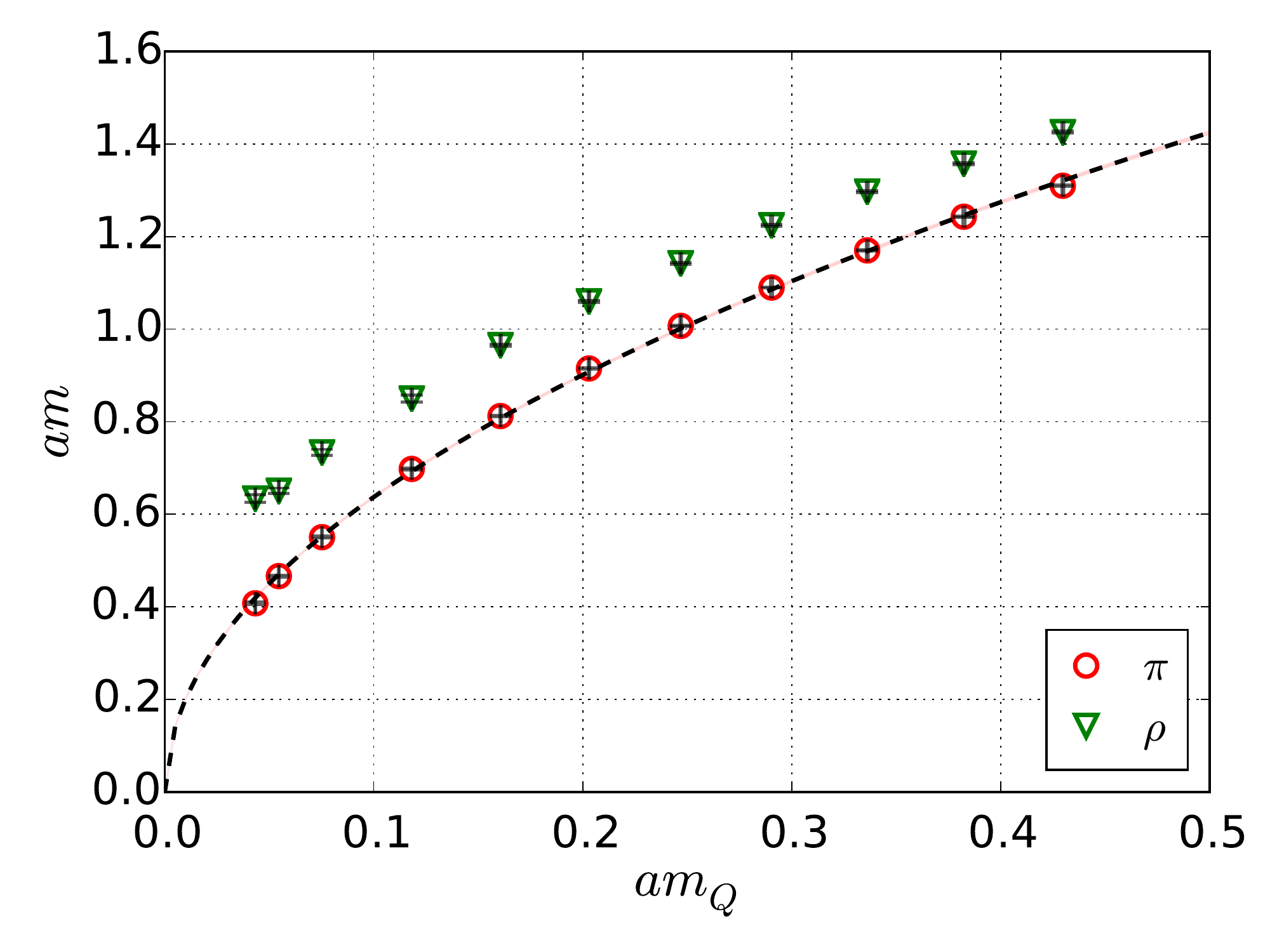}
\caption{Pseudoscalar and vector masses as a function of $m_Q$ for the two flavor theory.
The results correspond to lattices with $\beta_L=1.0$. The pseudoscalar mass approaches zero
in the chiral limit as $M_\pi\propto\sqrt{m_Q}$ (dashed curve), while the vector mass has a finite intercept.}
\label{Fig:nf2spectrum}
\end{figure}

Let us then turn to the details of the measurements. We start
with the spectrum of the two and four flavor theories.
At $N_f = 2$ we expect to see clear QCD-like chiral symmetry breaking behaviour.  In the analysis we use $\beta_L=1$, $L^3 \times T = 24^3\times 48$ lattices.  The pseudoscalar (``$\pi$'') and vector (``$\rho$'') spectrum is shown in Figure \ref{Fig:nf2spectrum}
and the expected behavior is observed: the pseudoscalar mass is compatible with
the behaviour $M_\pi\sim m_Q^{1/2}$ at small $m_Q$, while
the vector mass has a finite intercept as the quark mass approaches zero.
The ratio $m_\rho/m_\pi$ diverges as $m_Q\rightarrow 0$.

\begin{figure}
\includegraphics[width=.45\textwidth]{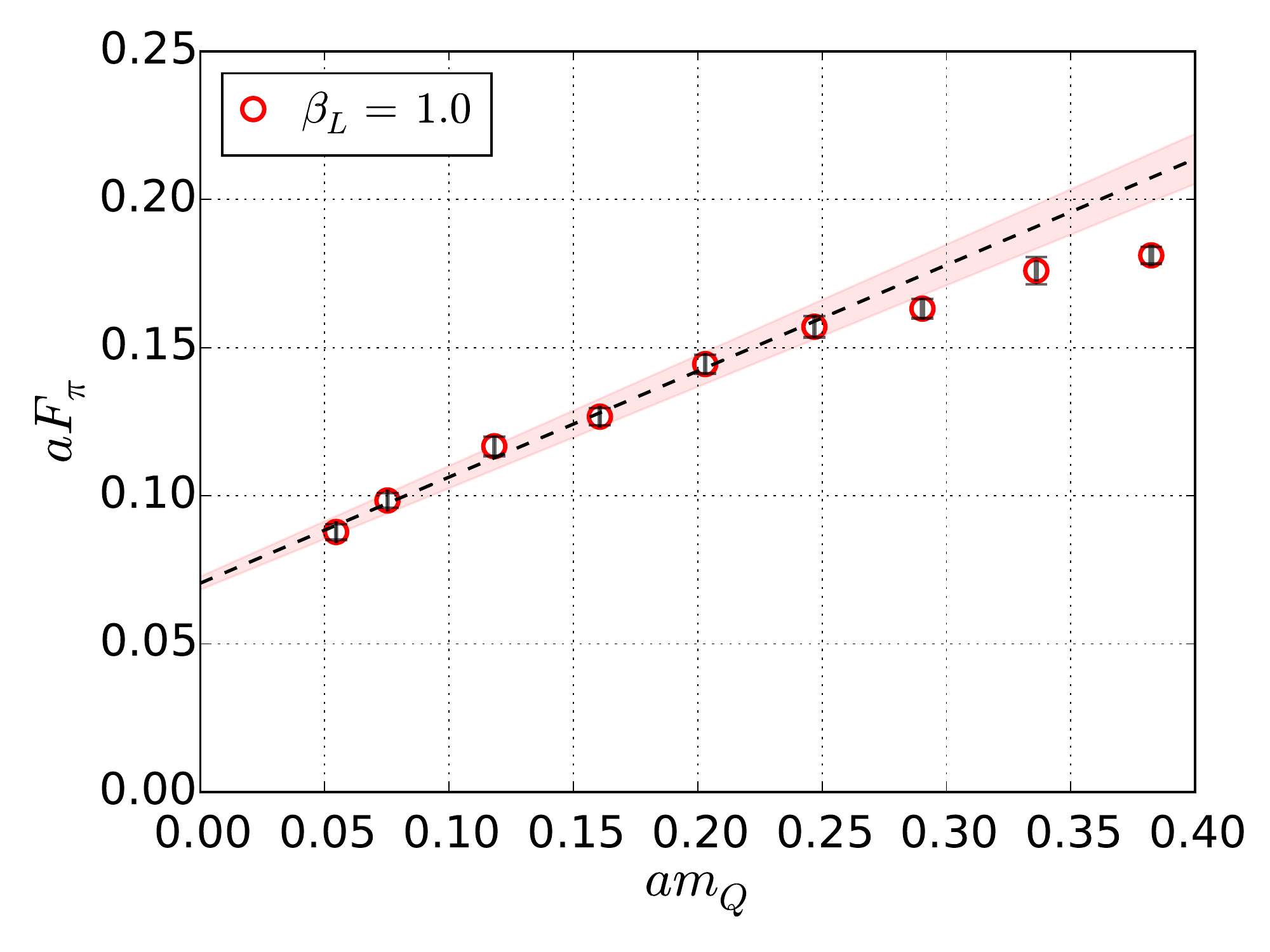}
\caption{Pseudoscalar decay constant as a function of $m_Q$ for the two flavor theory with $\beta_L=1.0$. The dashed line
is a linear fit of the form $y=\alpha+k x$ with $\alpha=0.070$ and $k=0.36$.}
\label{Fig:nf2decay}
\end{figure}

We have also measured the corresponding decay constants; the results for $F_\pi$ are shown in Figure \ref{Fig:nf2decay}.
Towards the chiral limit the data is
well described by a fit of the form
\begin{equation}
  aF_\pi(m_Q) = aF_{\pi,0} + k (am_Q),
  \label{fpifit}
\end{equation}
where $aF_{\pi,0}=0.070\pm 0.002$ and $k=0.358\pm 0.016$.
The result is consistent with the condition $F_\pi L > 1$ at $L=24a$, which
indicates that the finite volume effects should be under control in the chiral
extrapolation. In addition, $m_\pi L \gg 1$ for all of our measurements.
Because the behaviour at $N_f=2$ is uncontroversial, we do not repeat the
simulations at different lattice spacings (different $\beta_L$) and volumes.

The $N_f=4$ theory is expected to show similar behaviour.
However, because we are approaching the conformal window the pseudoscalar
decay constant is expected to be
smaller and thus more difficult to measure in the chiral limit.
In this case we use three different lattice couplings
$\beta_L=0.6$, $0.8$ and $1.0$ on lattice size $L^3\times T=24^3\times 48$, with
the addition of $32^3\times 60$ lattices at $\beta_L=0.8$ at small quark masses.

\begin{figure}
 \includegraphics[width=.45\textwidth]{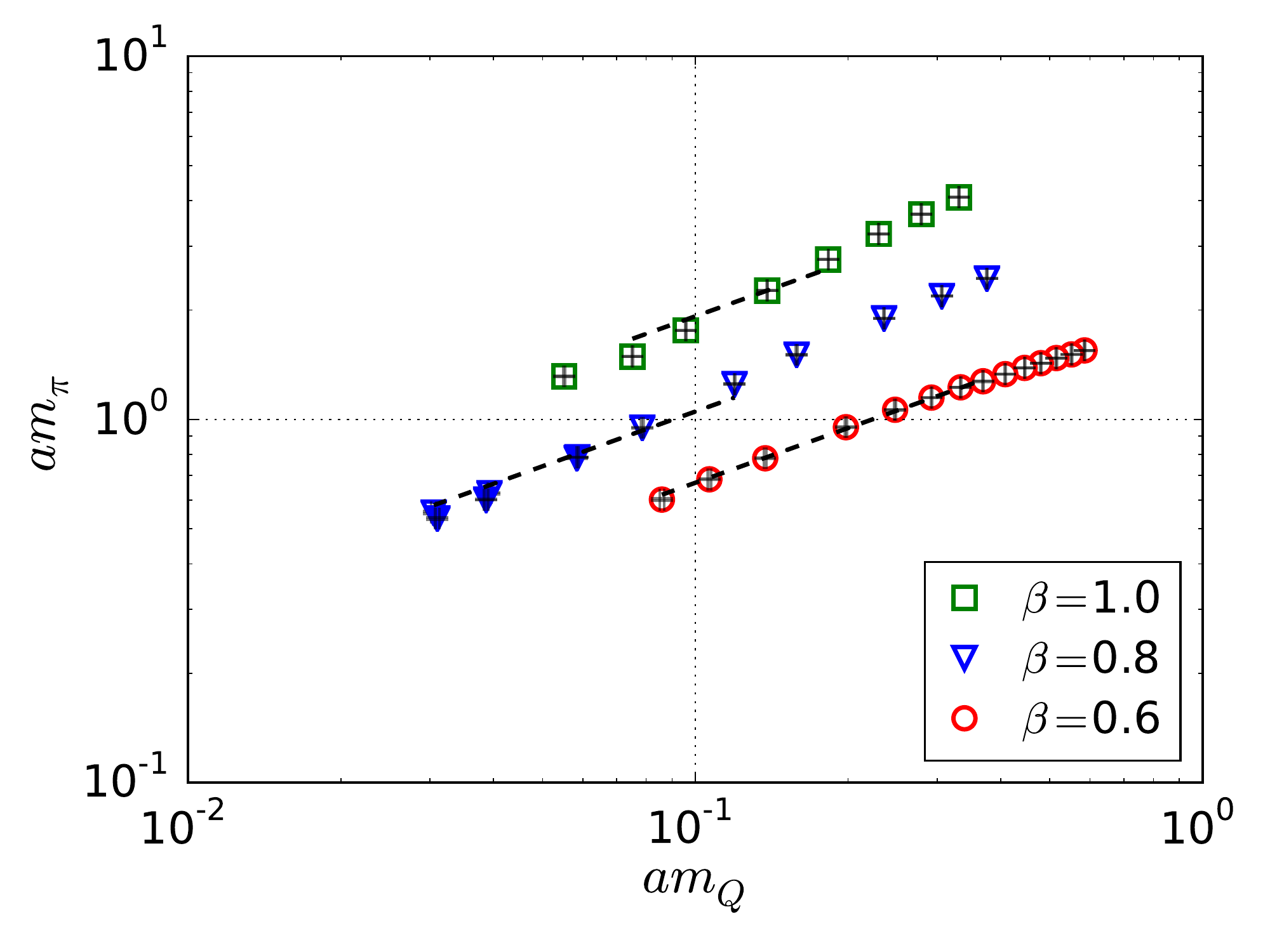}
 \caption{The pseudoscalar mass as a function of the quark mass in the four
 flavor theory at $\beta_L=0.6$, $0.8,$ and $1.0$. The pseudoscalar masses are multiplied by factors of $1$, $2$ and $4$, respectively. Empty markers refer to lattice size $L^3 \times T = 24^3\times 48$ and filled points to lattice size $L^3 \times T = 32^3\times 60$. Strong finite volume effects appear towards the chiral limit, but larger lattices allow for consistent results at smaller masses.}
 \label{Fig:nf4pion}
 \end{figure}

\begin{figure}
 \includegraphics[width=.45\textwidth]{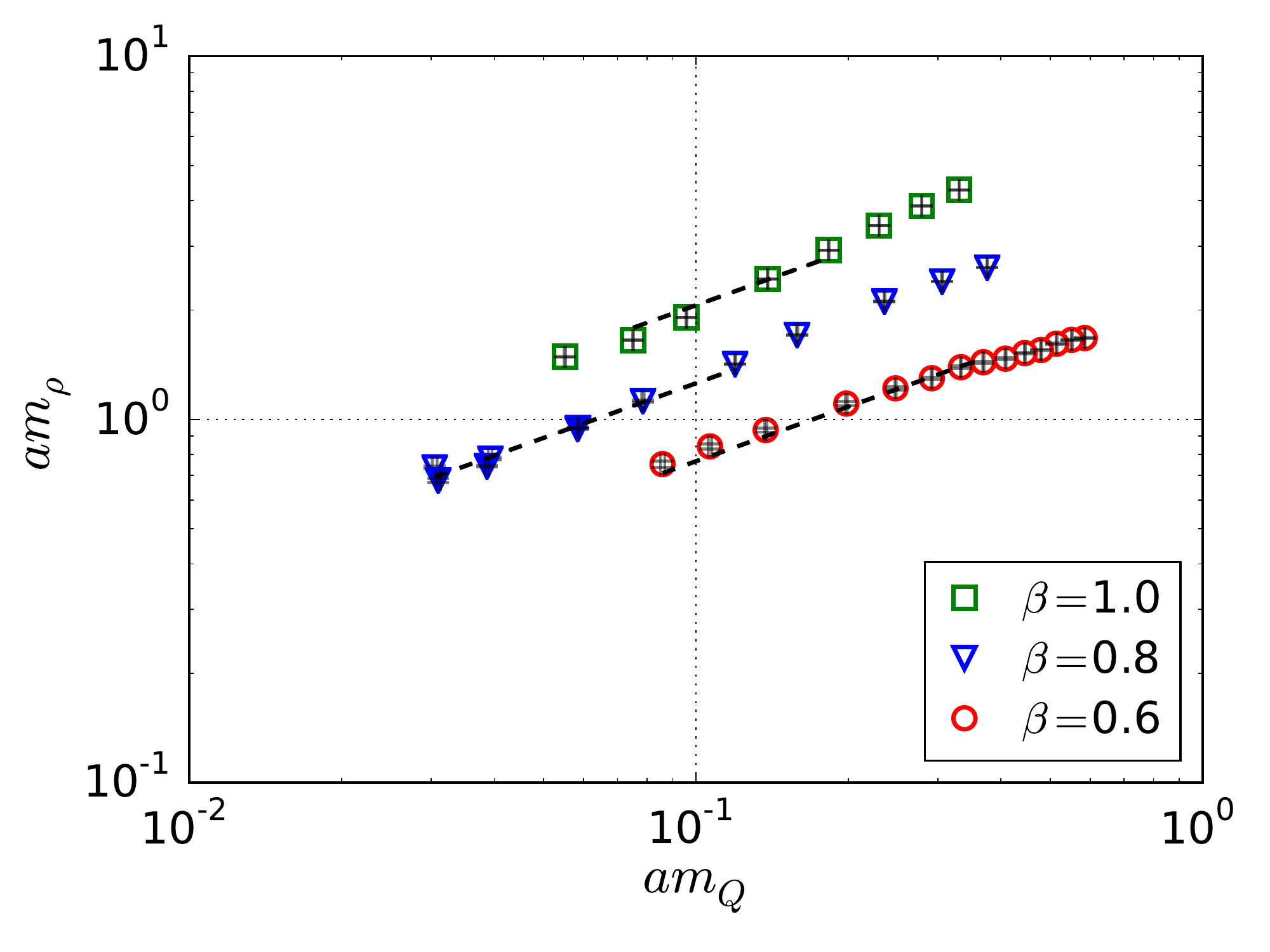}
 \caption{The vector mass as a function of the quark mass in the four
 flavor theory at $\beta_L=0.6$, $0.8,$ and $1.0$. The pseudoscalar masses are multiplied by factors of $1$, $2$ and $4$, respectively. Empty markers refer to lattice size $L^3 \times T = 24^3\times 48$ and filled points to lattice size $L^3 \times T = 32^3\times 60$. Strong finite volume effects appear towards the chiral limit, but larger lattices allow for consistent results at smaller masses.}
 \label{Fig:nf4rho}
 \end{figure}

\begin{figure}
\includegraphics[width=.45\textwidth]{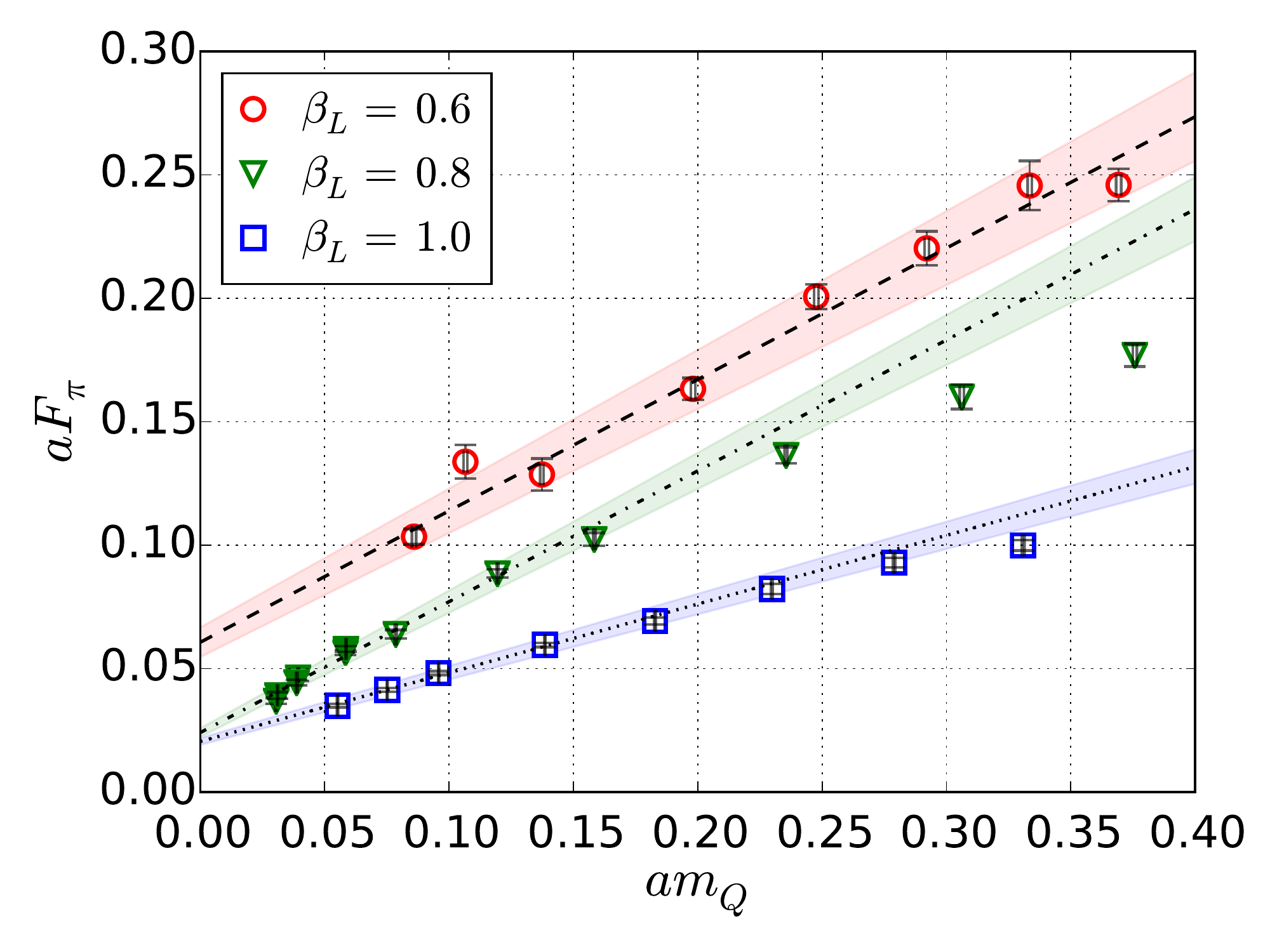}
\caption{The pseudoscalar decay constant as a function of $m_Q$ for the four flavor theory with $\beta_L=0.6,$ $0.8$ and $1.0$. Empty markers refer to lattice size $L^3 \times T = 24^3\times 48$ and filled points to lattice size $L^3 \times T = 32^3\times 60$.}
\label{Fig:nf4decay}
\end{figure}

The vector and pseudoscalar mass measurements are shown in Figures
\ref{Fig:nf4pion} and \ref{Fig:nf4rho}.
At the strongest coupling $\beta_L=0.6$ we observe a clear
$M_\pi\propto \sqrt{m_Q}$ behaviour, compatible with the chiral symmetry
breaking, whereas the square root fit becomes progressively less satisfactory
as $\beta_L$ increases.  This is likely caused by finite volume effects.  These
can be quantified by the pseudoscalar decay constant, shown in
Figure~\ref{Fig:nf4decay}.  Again a linear fit of the form in
Eq.~\eqref{fpifit},
works well, with the result
\begin{eqnarray}
  \beta_L &=& 0.6:\quad a F_{\pi,0}=0.061\pm 0.006,\quad k=0.532\pm 0.030, \nonumber \\
  \beta_L &=& 0.8:\quad a F_{\pi,0}=0.024\pm 0.002,\quad k=0.530\pm 0.028, \nonumber \\
  \beta_L &=& 1.0:\quad a F_{\pi,0}=0.020\pm0.001,\quad k=0.278\pm0.014.\nonumber
\end{eqnarray}
In this case $F_\pi L < 1$ at weaker couplings $\beta_L = 0.8$ and $1.0$, for both $L/a = 24$ and $32$. This means that the physical volume is probably not
large enough for a reliable chiral extrapolation,
possibly explaining the questionable square root fits for the pseudoscalar masses.  On the other hand, at $\beta_L = 0.6$ we have $F_\pi L \approx 1.6$.  Indeed, it is this constraint which forces us to use small $\beta_L$ (large lattice spacing) in simulations.
We remark that $M_\pi L \gg 1$ for all measurements, implying that the finite volume effects due to pions propagating around the lattice are small.  Thus, we conclude that the $N_f=4$ data
is consistent with chiral symmetry breaking, compatible with perturbative analysis and also
with the non-perturbative lattice coupling constant measurement in ref.~\cite{Karavirta:2011zg}.

%
%

\begin{figure}
 \includegraphics[width=.45\textwidth]{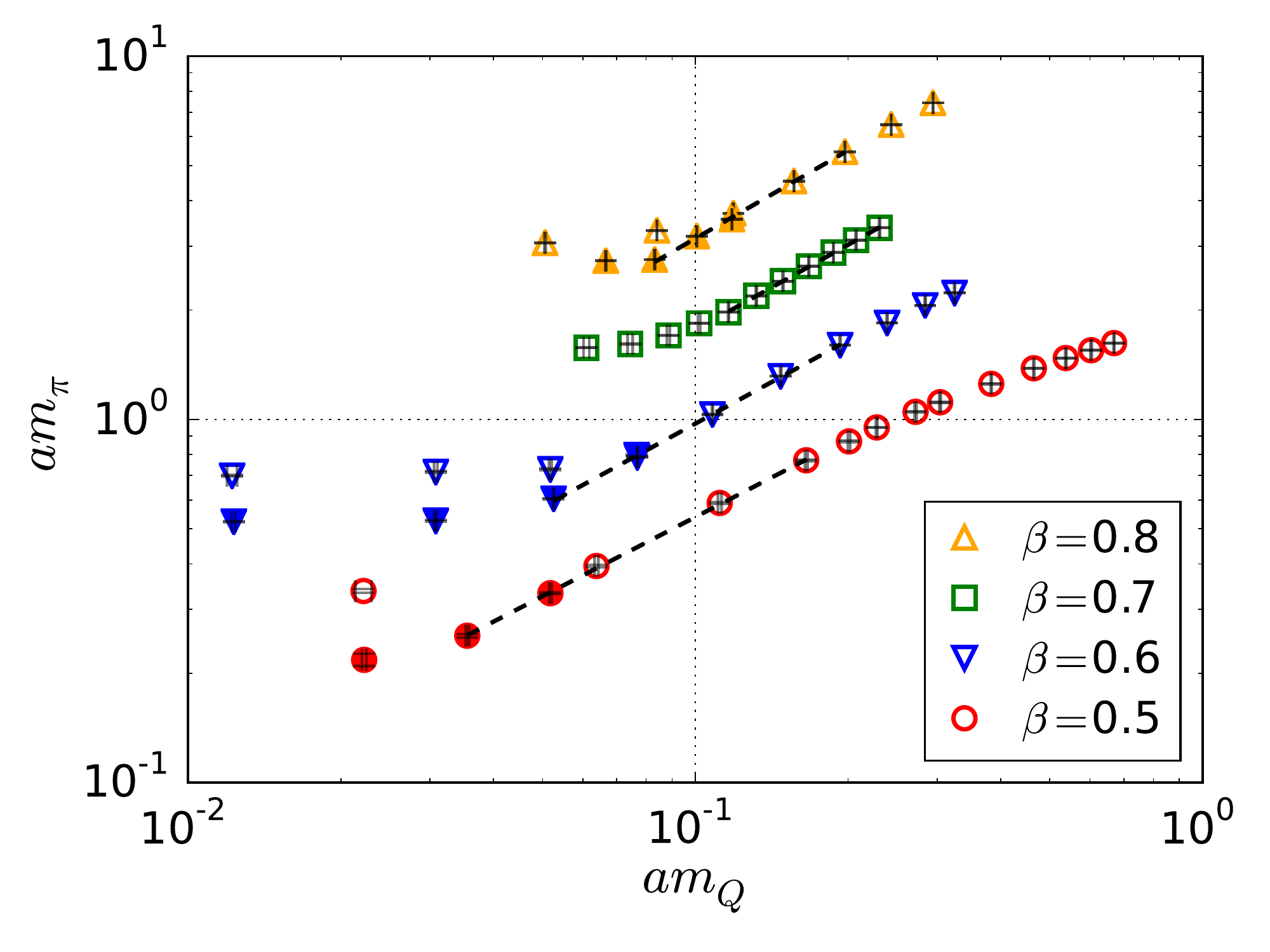}
 \caption{The pseudoscalar mass as a function of the quark mass in the six
 flavor theory at $\beta_L=0.5$, $0.6,$ $0.7$ and $0.8$. The pseudoscalar masses are multiplied by factors of $1$, $2$, $4$ and $8$, respectively. Empty markers refer to lattice size $L^3 \times T = 24^3\times 48$ and filled points to lattice size $L^3 \times T = 32^3\times 60$. Strong finite volume effects appear towards the chiral limit, but larger lattices allow for consistent results at smaller masses.}
 \label{Fig:nf6pion}
 \end{figure}

\begin{figure}
 \includegraphics[width=.45\textwidth]{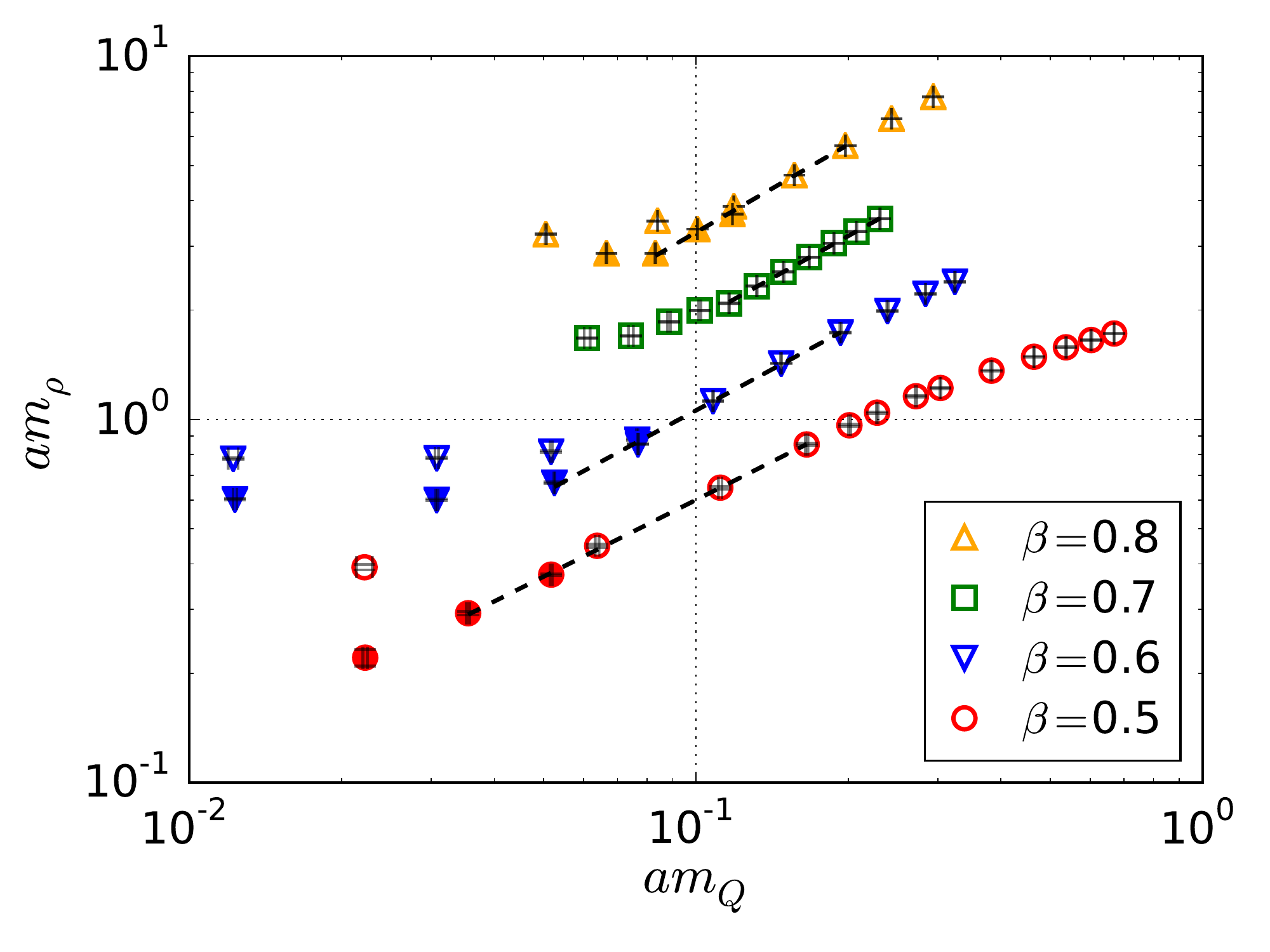}
 \caption{The vector mass as a function of the quark mass in the six
 flavor theory at $\beta_L=0.6,$, $0.7$, $0.8$ and $1.0$. The vector masses are multiplied by factors of $1$, $2$, $4$ and $8$, respectively. Empty markers refer to lattice size $L^3 \times T = 24^3\times 48$ and filled points to lattice size $L^3 \times T = 32^3\times 60$.}
 \label{Fig:nf6rho}
 \end{figure}

Let us then turn to the more interesting case of the six flavor theory.
Here we observe strong volume dependence on the pseudoscalar and vector meson masses at small quark masses, as illustrated
in Figures~\ref{Fig:nf6pion} and~\ref{Fig:nf6rho}, respectively,
measured using $\beta_L=0.5$, $0.6$ and $0.8$.  This is especially clearly visible in the data at
$\beta_L=0.6$: meson masses saturate or start even increasing as $m_Q$ becomes
too small. The volume dependence is clearly observed as this levelling off point moves to smaller $m_Q$ as the volume increases.


As shown in the figures, larger volumes allow one to run towards smaller masses, and extrapolating from these results,
we conclude that both the vector and the pseudoscalar scale towards zero with quark mass. This behavior is consistent with the existence of an infrared fixed point.

\begin{figure}
\includegraphics[width=.45\textwidth]{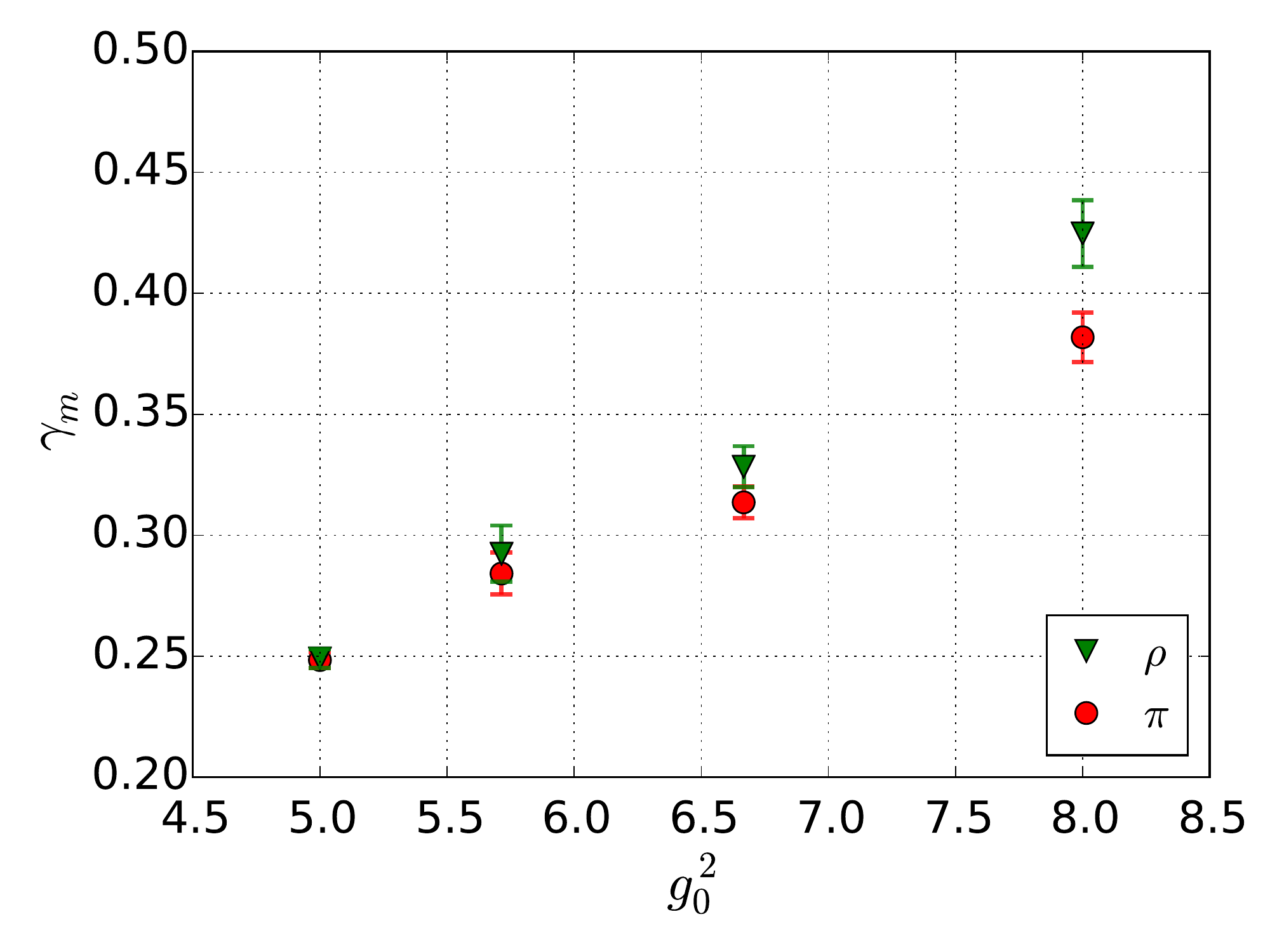}
\caption{The anomalous dimension as a function of the bare coupling $\beta_L$  as determined from the spectrum of the six flavor theory.}
\label{Fig:nf6gamma}
\end{figure}

We can determine the anomalous dimension of the quark mass
operator from the spectrum: the scaling of the hadron masses \kr{at the
  infrared fixed point is $M_h\sim m_q^{1/(1+\gamma(g^\ast))}$, where $g^\ast$ is the fixed point coupling.  On a lattice, this is the expected behaviour 
in the limit $m_q\rightarrow 0$ and $L\rightarrow\infty$, independent of the bare lattice gauge coupling $g_0^2$.  However, in practice it is not possible to
reach small enough $m_q$ and large enough $L$ in order to observe this, unless the bare coupling tuned so that the coupling on the lattice scale is close to $g^\ast$.  Instead, we expect that the hadron masses behave approximately as
$
M_h \sim m_q^{1/(1+\gamma(g(\Lambda)))}
$,
where the coupling is to be evaluated at the scale $\Lambda \sim 1/L$ or $1/M_h$.
Because the coupling evolves very slowly near the fixed point and we have
only one or two lattice sizes for each bare coupling $\beta_L = 4/g_0^2$, we expect that $M_h$ obeys nearly pure power law behaviour with constant
$\gamma(g(1/L))$.  This is indeed
the case, as can be seen in Figures~\ref{Fig:nf6pion} and~\ref{Fig:nf6rho},
where power law fits at each $\beta_L$ are shown, with the resulting
$\gamma$-values shown in Figure \ref{Fig:nf6gamma}.}

\kr{Let us now relate the above result to 
physical couplings measured at the lattice scale $1/L$.  Here we can use the
recent results in ref.~\cite{Leino:2017hgm}, where the running coupling of SU(2) with $N_f=6$ fermions has been measured
using the Schrödinger functional method in combination with gradient flow.
Because the lattice action used in this paper was exactly the same as here,
we can map our $(\beta_L,L)$ combinations to the tabulated gradient flow scheme
couplings $g_{\rm{GF}}$, using the preferred scheme parameters of
ref.~\cite{Leino:2017hgm}.}

\begin{figure}
\includegraphics[width=.45\textwidth]{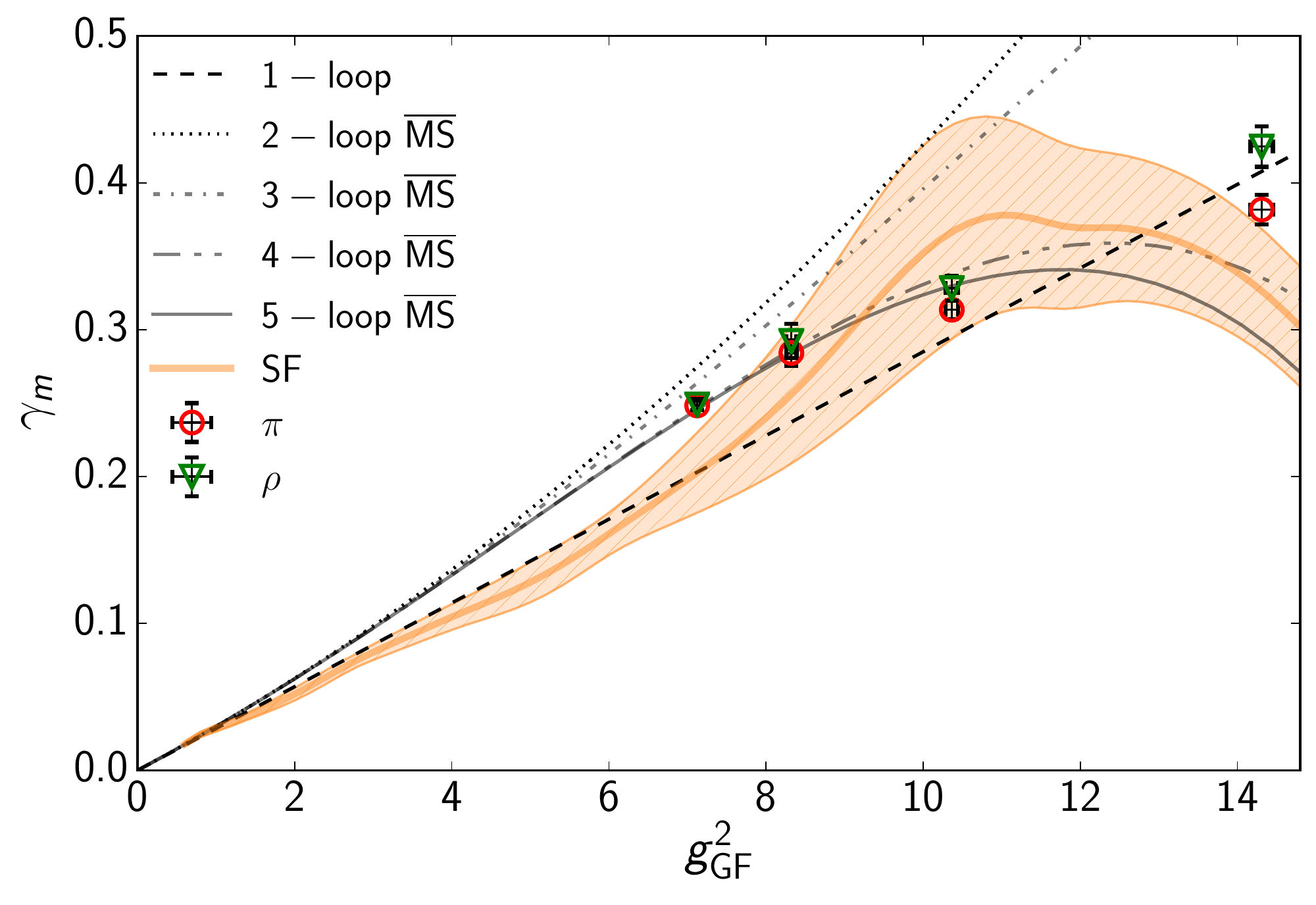}
\caption{The anomalous dimension as a function of the running coupling $g_{\rm{GF}}$ measured using the gradient flow method. The shaded band is the
result from~\cite{Leino:2017hgm} obtained using the step-scaling method. The
open symbols show the results obtained in this work from the pseudoscalar and vector meson spectra. The curves show perturbative results at different orders of perturbation theory.}
\label{Fig:gammat}
\end{figure}

In Figure~\ref{Fig:gammat} we show the values of the anomalous dimension determined from the spectra of pseudoscalar and vector mesons together with the
result from~\cite{Leino:2017hgm}. The results of the present work are shown
with open circles and triangles corresponding, respectively, to the
pseudoscalar and vector mesons, while the earlier step-scaling result is shown
by the shaded band. Also shown in the figure are the results from perturbation
theory at different orders~\cite{Luthe:2016xec}.
\kr{The observed very good agreement between the high-order perturbative result and the measurement from the hadron spectrum is partly coincidental,
  due to different schemes.  However, the  match between the step scaling and the hadron  spectrum results does not suffer from scheme dependence and
  indicate that 
  different nonperturbative methods yield results consistent with each other.}  
At the fixed point $g^\ast\simeq 14.5$, \kr{determined in~\cite{Leino:2017hgm}}, the step
scaling method gives $\gamma(g^*)=0.283(2)\pm 0.01$ with systematic
and statistical errors.

\begin{figure}
\includegraphics[width=.45\textwidth]{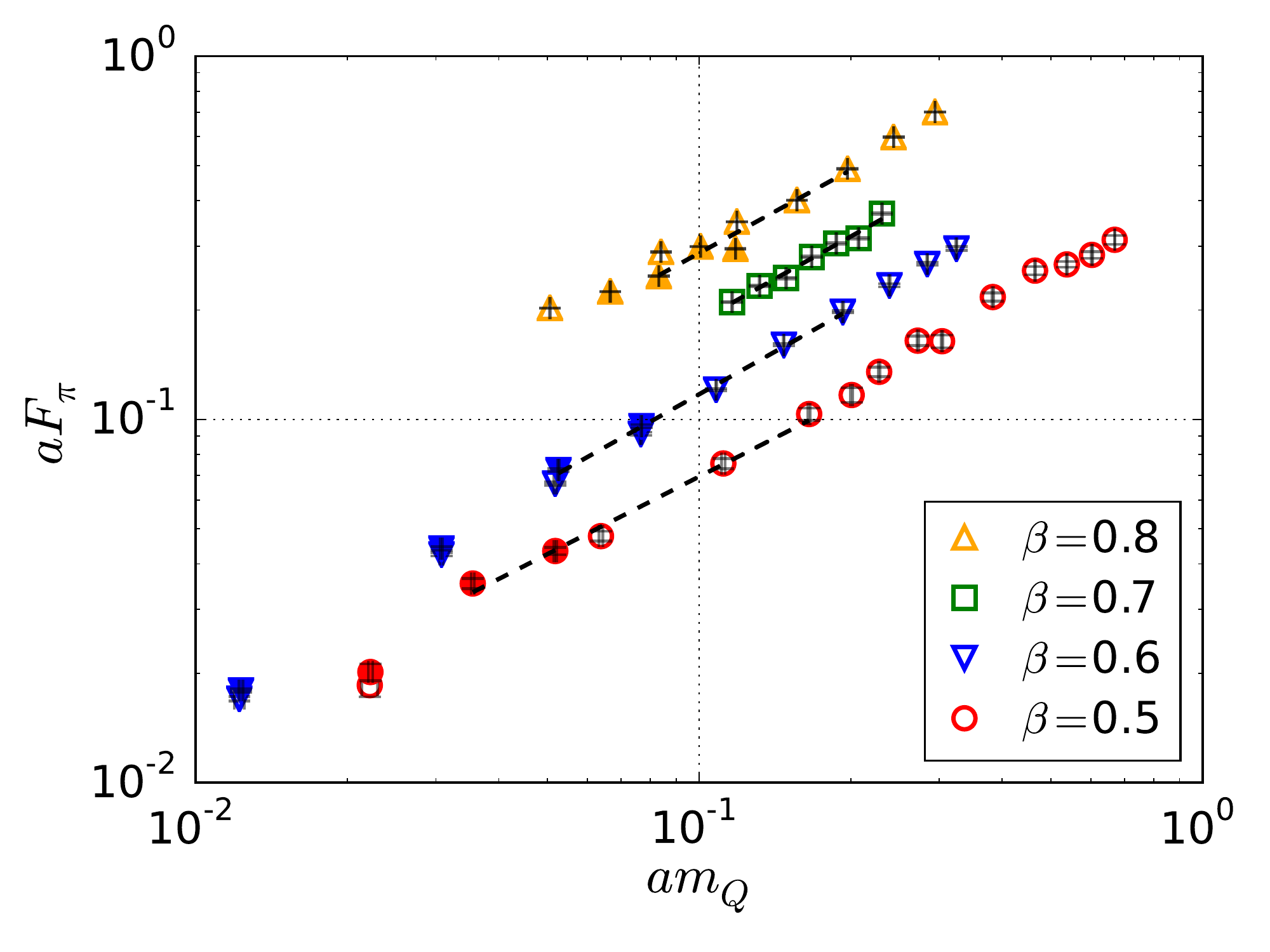}
\caption{The pseudoscalar decay constant as a function of $m_Q$ for the six flavor theory corresponding to lattices $\beta_L=0.5$, $0.6$ and $0.8$. The pseudoscalar decay constants are multiplied by factors of $1$, $2$, $4$ and $8$, respectively. Empty markers refer to lattice size $L^3 \times T = 24^3\times 48$ and filled points to lattice size $L^3 \times T = 32^3\times 60$.
}
\label{Fig:nf6decay}
\end{figure}

On these same lattices we have again measured also the decay constant of the pseudoscalar meson
in the six flavor theory. The results are shown in Figure~\ref{Fig:nf6decay} for
$\beta_L=0.5$, $0.6$, $0.7$ and $0.8$.
The pseudoscalar decay constant is observed to scale towards zero
in the chiral limit which lends further support to interpreting the
results of the spectral measurement as arising due to infrared conformality
of the six flavor theory. The data can be fitted with a power law, $aF_\pi=\alpha (am_Q)^b$ with the exponent $b$ taking similar values as in the mass spectrum case.

\section{Conclusions}
\label{sec:chekout}
In this paper we have studied the vacuum spectrum of SU(2) gauge theory with
various numbers of Dirac fermions in the fundamental representation of the
gauge group. We focused on the properties of the spectra as a function of quark
mass.

For a small numbers of flavors, $N_f=2$ and 4 we observed behavior consistent
with the expectations based on a QCD-like theory: chiral symmetry is
spontaneously broken in the vacuum leading to the usual $m_q$ dependence of the pseudoscalar and vector meson masses.
Different behavior was observed for the six flavor theory as we found
behavior consistent with the existence of an infrared fixed point in this theory. All meson masses were shown to scale towards zero as quark mass is decreased. The approach to the chiral limit in the spectra showed strong volume dependence and we demonstrated how larger lattices allow simulations at
smaller quark masses.

We then extracted the anomalous dimension of the mass operator from the scaling
of meson masses. Using the results of~\cite{Leino:2017hgm} we were able to
compare the result from the measurement of the spectra to the one obtained
using step-scaling analysis. We found that the results of the present paper are
fully consistent with the earlier work of~\cite{Leino:2017hgm}. The main consequence of our work therefore is to
strengthen the conclusions of~\cite{Leino:2017hgm} by providing
more evidence for the existence of a fixed point in the SU(2) gauge theory with
six flavors.

\FloatBarrier
\acknowledgments{
This work is supported by the Academy of Finland grants 310130, 308791 and 267286.
S.T. is supported by the Magnus Ehrnrooth foundation 
and V.L by the Jenny and Antti Wihuri foundation.
The simulations were performed at the Finnish IT Center for Science (CSC), Espoo, Finland.
}

\bibliography{su2_nfdep}{}

\begin{thebibliography}{21}%
\makeatletter
\providecommand \@ifxundefined [1]{%
 \@ifx{#1\undefined}
}%
\providecommand \@ifnum [1]{%
 \ifnum #1\expandafter \@firstoftwo
 \else \expandafter \@secondoftwo
 \fi
}%
\providecommand \@ifx [1]{%
 \ifx #1\expandafter \@firstoftwo
 \else \expandafter \@secondoftwo
 \fi
}%
\providecommand \natexlab [1]{#1}%
\providecommand \enquote  [1]{``#1''}%
\providecommand \bibnamefont  [1]{#1}%
\providecommand \bibfnamefont [1]{#1}%
\providecommand \citenamefont [1]{#1}%
\providecommand \href@noop [0]{\@secondoftwo}%
\providecommand \href [0]{\begingroup \@sanitize@url \@href}%
\providecommand \@href[1]{\@@startlink{#1}\@@href}%
\providecommand \@@href[1]{\endgroup#1\@@endlink}%
\providecommand \@sanitize@url [0]{\catcode `\\12\catcode `\$12\catcode
  `\&12\catcode `\#12\catcode `\^12\catcode `\_12\catcode `\%12\relax}%
\providecommand \@@startlink[1]{}%
\providecommand \@@endlink[0]{}%
\providecommand \url  [0]{\begingroup\@sanitize@url \@url }%
\providecommand \@url [1]{\endgroup\@href {#1}{\urlprefix }}%
\providecommand \urlprefix  [0]{URL }%
\providecommand \Eprint [0]{\href }%
\providecommand \doibase [0]{http://dx.doi.org/}%
\providecommand \selectlanguage [0]{\@gobble}%
\providecommand \bibinfo  [0]{\@secondoftwo}%
\providecommand \bibfield  [0]{\@secondoftwo}%
\providecommand \translation [1]{[#1]}%
\providecommand \BibitemOpen [0]{}%
\providecommand \bibitemStop [0]{}%
\providecommand \bibitemNoStop [0]{.\EOS\space}%
\providecommand \EOS [0]{\spacefactor3000\relax}%
\providecommand \BibitemShut  [1]{\csname bibitem#1\endcsname}%
\let\auto@bib@innerbib\@empty
\bibitem [{\citenamefont {Susskind}(1979)}]{Susskind:1978ms}%
  \BibitemOpen
  \bibfield  {author} {\bibinfo {author} {\bibfnamefont {L.}~\bibnamefont
  {Susskind}},\ }\href {\doibase 10.1103/PhysRevD.20.2619} {\bibfield
  {journal} {\bibinfo  {journal} {Phys. Rev.}\ }\textbf {\bibinfo {volume}
  {D20}},\ \bibinfo {pages} {2619} (\bibinfo {year} {1979})}\BibitemShut
  {NoStop}%
\bibitem [{\citenamefont {Weinberg}(1979)}]{Weinberg:1979bn}%
  \BibitemOpen
  \bibfield  {author} {\bibinfo {author} {\bibfnamefont {S.}~\bibnamefont
  {Weinberg}},\ }\href {\doibase 10.1103/PhysRevD.19.1277} {\bibfield
  {journal} {\bibinfo  {journal} {Phys. Rev.}\ }\textbf {\bibinfo {volume}
  {D19}},\ \bibinfo {pages} {1277} (\bibinfo {year} {1979})}\BibitemShut
  {NoStop}%
\bibitem [{\citenamefont {Hill}\ and\ \citenamefont
  {Simmons}(2003)}]{Hill:2002ap}%
  \BibitemOpen
  \bibfield  {author} {\bibinfo {author} {\bibfnamefont {C.~T.}\ \bibnamefont
  {Hill}}\ and\ \bibinfo {author} {\bibfnamefont {E.~H.}\ \bibnamefont
  {Simmons}},\ }\href {\doibase 10.1016/S0370-1573(03)00140-6} {\bibfield
  {journal} {\bibinfo  {journal} {Phys. Rept.}\ }\textbf {\bibinfo {volume}
  {381}},\ \bibinfo {pages} {235} (\bibinfo {year} {2003})},\ \bibinfo {note}
  {[Erratum: Phys. Rept.390,553(2004)]},\ \Eprint
  {http://arxiv.org/abs/hep-ph/0203079} {arXiv:hep-ph/0203079 [hep-ph]}
  \BibitemShut {NoStop}%
\bibitem [{\citenamefont {Sannino}(2008)}]{Sannino:2008ha}%
  \BibitemOpen
  \bibfield  {author} {\bibinfo {author} {\bibfnamefont {F.}~\bibnamefont
  {Sannino}},\ }\href@noop {} {\  (\bibinfo {year} {2008})},\ \Eprint
  {http://arxiv.org/abs/0804.0182} {arXiv:0804.0182 [hep-ph]} \BibitemShut
  {NoStop}%
\bibitem [{\citenamefont {Sannino}\ and\ \citenamefont
  {Tuominen}(2005)}]{Sannino:2004qp}%
  \BibitemOpen
  \bibfield  {author} {\bibinfo {author} {\bibfnamefont {F.}~\bibnamefont
  {Sannino}}\ and\ \bibinfo {author} {\bibfnamefont {K.}~\bibnamefont
  {Tuominen}},\ }\href {\doibase 10.1103/PhysRevD.71.051901} {\bibfield
  {journal} {\bibinfo  {journal} {Phys. Rev.}\ }\textbf {\bibinfo {volume}
  {D71}},\ \bibinfo {pages} {051901} (\bibinfo {year} {2005})},\ \Eprint
  {http://arxiv.org/abs/hep-ph/0405209} {arXiv:hep-ph/0405209 [hep-ph]}
  \BibitemShut {NoStop}%
\bibitem [{\citenamefont {Banks}\ and\ \citenamefont
  {Zaks}(1982)}]{Banks:1981nn}%
  \BibitemOpen
  \bibfield  {author} {\bibinfo {author} {\bibfnamefont {T.}~\bibnamefont
  {Banks}}\ and\ \bibinfo {author} {\bibfnamefont {A.}~\bibnamefont {Zaks}},\
  }\href {\doibase 10.1016/0550-3213(82)90035-9} {\bibfield  {journal}
  {\bibinfo  {journal} {Nucl. Phys.}\ }\textbf {\bibinfo {volume} {B196}},\
  \bibinfo {pages} {189} (\bibinfo {year} {1982})}\BibitemShut {NoStop}%
\bibitem [{\citenamefont {Ohki}\ \emph {et~al.}(2010)\citenamefont {Ohki},
  \citenamefont {Aoyama}, \citenamefont {Itou}, \citenamefont {Kurachi},
  \citenamefont {Lin}, \citenamefont {Matsufuru}, \citenamefont {Onogi},
  \citenamefont {Shintani},\ and\ \citenamefont {Yamazaki}}]{Ohki:2010sr}%
  \BibitemOpen
  \bibfield  {author} {\bibinfo {author} {\bibfnamefont {H.}~\bibnamefont
  {Ohki}}, \bibinfo {author} {\bibfnamefont {T.}~\bibnamefont {Aoyama}},
  \bibinfo {author} {\bibfnamefont {E.}~\bibnamefont {Itou}}, \bibinfo {author}
  {\bibfnamefont {M.}~\bibnamefont {Kurachi}}, \bibinfo {author} {\bibfnamefont
  {C.~J.~D.}\ \bibnamefont {Lin}}, \bibinfo {author} {\bibfnamefont
  {H.}~\bibnamefont {Matsufuru}}, \bibinfo {author} {\bibfnamefont
  {T.}~\bibnamefont {Onogi}}, \bibinfo {author} {\bibfnamefont
  {E.}~\bibnamefont {Shintani}}, \ and\ \bibinfo {author} {\bibfnamefont
  {T.}~\bibnamefont {Yamazaki}},\ }\bibfield  {booktitle} {\emph {\bibinfo
  {booktitle} {{Proceedings, 28th International Symposium on Lattice field
  theory (Lattice 2010)}}},\ }\href@noop {} {\bibfield  {journal} {\bibinfo
  {journal} {PoS}\ }\textbf {\bibinfo {volume} {LATTICE2010}},\ \bibinfo
  {pages} {066} (\bibinfo {year} {2010})},\ \Eprint
  {http://arxiv.org/abs/1011.0373} {arXiv:1011.0373 [hep-lat]} \BibitemShut
  {NoStop}%
\bibitem [{\citenamefont {Bursa}\ \emph {et~al.}(2011)\citenamefont {Bursa},
  \citenamefont {Del~Debbio}, \citenamefont {Keegan}, \citenamefont {Pica},\
  and\ \citenamefont {Pickup}}]{Bursa:2010xn}%
  \BibitemOpen
  \bibfield  {author} {\bibinfo {author} {\bibfnamefont {F.}~\bibnamefont
  {Bursa}}, \bibinfo {author} {\bibfnamefont {L.}~\bibnamefont {Del~Debbio}},
  \bibinfo {author} {\bibfnamefont {L.}~\bibnamefont {Keegan}}, \bibinfo
  {author} {\bibfnamefont {C.}~\bibnamefont {Pica}}, \ and\ \bibinfo {author}
  {\bibfnamefont {T.}~\bibnamefont {Pickup}},\ }\href {\doibase
  10.1016/j.physletb.2010.12.050} {\bibfield  {journal} {\bibinfo  {journal}
  {Phys. Lett.}\ }\textbf {\bibinfo {volume} {B696}},\ \bibinfo {pages} {374}
  (\bibinfo {year} {2011})},\ \Eprint {http://arxiv.org/abs/1007.3067}
  {arXiv:1007.3067 [hep-ph]} \BibitemShut {NoStop}%
\bibitem [{\citenamefont {Karavirta}\ \emph {et~al.}(2012)\citenamefont
  {Karavirta}, \citenamefont {Rantaharju}, \citenamefont {Rummukainen},\ and\
  \citenamefont {Tuominen}}]{Karavirta:2011zg}%
  \BibitemOpen
  \bibfield  {author} {\bibinfo {author} {\bibfnamefont {T.}~\bibnamefont
  {Karavirta}}, \bibinfo {author} {\bibfnamefont {J.}~\bibnamefont
  {Rantaharju}}, \bibinfo {author} {\bibfnamefont {K.}~\bibnamefont
  {Rummukainen}}, \ and\ \bibinfo {author} {\bibfnamefont {K.}~\bibnamefont
  {Tuominen}},\ }\href {\doibase 10.1007/JHEP05(2012)003} {\bibfield  {journal}
  {\bibinfo  {journal} {JHEP}\ }\textbf {\bibinfo {volume} {05}},\ \bibinfo
  {pages} {003} (\bibinfo {year} {2012})},\ \Eprint
  {http://arxiv.org/abs/1111.4104} {arXiv:1111.4104 [hep-lat]} \BibitemShut
  {NoStop}%
\bibitem [{\citenamefont {Lewis}\ \emph {et~al.}(2012)\citenamefont {Lewis},
  \citenamefont {Pica},\ and\ \citenamefont {Sannino}}]{Lewis:2011zb}%
  \BibitemOpen
  \bibfield  {author} {\bibinfo {author} {\bibfnamefont {R.}~\bibnamefont
  {Lewis}}, \bibinfo {author} {\bibfnamefont {C.}~\bibnamefont {Pica}}, \ and\
  \bibinfo {author} {\bibfnamefont {F.}~\bibnamefont {Sannino}},\ }\href
  {\doibase 10.1103/PhysRevD.85.014504} {\bibfield  {journal} {\bibinfo
  {journal} {Phys. Rev.}\ }\textbf {\bibinfo {volume} {D85}},\ \bibinfo {pages}
  {014504} (\bibinfo {year} {2012})},\ \Eprint {http://arxiv.org/abs/1109.3513}
  {arXiv:1109.3513 [hep-ph]} \BibitemShut {NoStop}%
\bibitem [{\citenamefont {Hayakawa}\ \emph {et~al.}(2013)\citenamefont
  {Hayakawa}, \citenamefont {Ishikawa}, \citenamefont {Takeda},\ and\
  \citenamefont {Yamada}}]{Hayakawa:2013yfa}%
  \BibitemOpen
  \bibfield  {author} {\bibinfo {author} {\bibfnamefont {M.}~\bibnamefont
  {Hayakawa}}, \bibinfo {author} {\bibfnamefont {K.~I.}\ \bibnamefont
  {Ishikawa}}, \bibinfo {author} {\bibfnamefont {S.}~\bibnamefont {Takeda}}, \
  and\ \bibinfo {author} {\bibfnamefont {N.}~\bibnamefont {Yamada}},\ }\href
  {\doibase 10.1103/PhysRevD.88.094504} {\bibfield  {journal} {\bibinfo
  {journal} {Phys. Rev.}\ }\textbf {\bibinfo {volume} {D88}},\ \bibinfo {pages}
  {094504} (\bibinfo {year} {2013})},\ \Eprint {http://arxiv.org/abs/1307.6997}
  {arXiv:1307.6997 [hep-lat]} \BibitemShut {NoStop}%
\bibitem [{\citenamefont {Appelquist}\ \emph {et~al.}(2014)\citenamefont
  {Appelquist}, \citenamefont {Brower}, \citenamefont {Buchoff}, \citenamefont
  {Cheng}, \citenamefont {Fleming}, \citenamefont {Kiskis}, \citenamefont
  {Lin}, \citenamefont {Neil}, \citenamefont {Osborn}, \citenamefont {Rebbi}
  \emph {et~al.}}]{Appelquist:2013pqa}%
  \BibitemOpen
  \bibfield  {author} {\bibinfo {author} {\bibfnamefont {T.}~\bibnamefont
  {Appelquist}}, \bibinfo {author} {\bibfnamefont {R.}~\bibnamefont {Brower}},
  \bibinfo {author} {\bibfnamefont {M.}~\bibnamefont {Buchoff}}, \bibinfo
  {author} {\bibfnamefont {M.}~\bibnamefont {Cheng}}, \bibinfo {author}
  {\bibfnamefont {G.}~\bibnamefont {Fleming}}, \bibinfo {author} {\bibfnamefont
  {J.}~\bibnamefont {Kiskis}}, \bibinfo {author} {\bibfnamefont
  {M.}~\bibnamefont {Lin}}, \bibinfo {author} {\bibfnamefont {E.}~\bibnamefont
  {Neil}}, \bibinfo {author} {\bibfnamefont {J.}~\bibnamefont {Osborn}},
  \bibinfo {author} {\bibfnamefont {C.}~\bibnamefont {Rebbi}},  \emph
  {et~al.},\ }\href {\doibase 10.1103/PhysRevLett.112.111601} {\bibfield
  {journal} {\bibinfo  {journal} {Phys. Rev. Lett.}\ }\textbf {\bibinfo
  {volume} {112}},\ \bibinfo {pages} {111601} (\bibinfo {year} {2014})},\
  \Eprint {http://arxiv.org/abs/1311.4889} {arXiv:1311.4889 [hep-ph]}
  \BibitemShut {NoStop}%
\bibitem [{\citenamefont {Hietanen}\ \emph {et~al.}(2014)\citenamefont
  {Hietanen}, \citenamefont {Lewis}, \citenamefont {Pica},\ and\ \citenamefont
  {Sannino}}]{Hietanen:2014xca}%
  \BibitemOpen
  \bibfield  {author} {\bibinfo {author} {\bibfnamefont {A.}~\bibnamefont
  {Hietanen}}, \bibinfo {author} {\bibfnamefont {R.}~\bibnamefont {Lewis}},
  \bibinfo {author} {\bibfnamefont {C.}~\bibnamefont {Pica}}, \ and\ \bibinfo
  {author} {\bibfnamefont {F.}~\bibnamefont {Sannino}},\ }\href {\doibase
  10.1007/JHEP07(2014)116} {\bibfield  {journal} {\bibinfo  {journal} {JHEP}\
  }\textbf {\bibinfo {volume} {07}},\ \bibinfo {pages} {116} (\bibinfo {year}
  {2014})},\ \Eprint {http://arxiv.org/abs/1404.2794} {arXiv:1404.2794
  [hep-lat]} \BibitemShut {NoStop}%
\bibitem [{\citenamefont {Leino}\ \emph {et~al.}(2017)\citenamefont {Leino},
  \citenamefont {Rantaharju}, \citenamefont {Rantalaiho}, \citenamefont
  {Rummukainen}, \citenamefont {Suorsa},\ and\ \citenamefont
  {Tuominen}}]{Leino:2017lpc}%
  \BibitemOpen
  \bibfield  {author} {\bibinfo {author} {\bibfnamefont {V.}~\bibnamefont
  {Leino}}, \bibinfo {author} {\bibfnamefont {J.}~\bibnamefont {Rantaharju}},
  \bibinfo {author} {\bibfnamefont {T.}~\bibnamefont {Rantalaiho}}, \bibinfo
  {author} {\bibfnamefont {K.}~\bibnamefont {Rummukainen}}, \bibinfo {author}
  {\bibfnamefont {J.~M.}\ \bibnamefont {Suorsa}}, \ and\ \bibinfo {author}
  {\bibfnamefont {K.}~\bibnamefont {Tuominen}},\ }\href {\doibase
  10.1103/PhysRevD.95.114516} {\bibfield  {journal} {\bibinfo  {journal} {Phys.
  Rev.}\ }\textbf {\bibinfo {volume} {D95}},\ \bibinfo {pages} {114516}
  (\bibinfo {year} {2017})},\ \Eprint {http://arxiv.org/abs/1701.04666}
  {arXiv:1701.04666 [hep-lat]} \BibitemShut {NoStop}%
\bibitem [{\citenamefont {Leino}\ \emph
  {et~al.}(2018{\natexlab{a}})\citenamefont {Leino}, \citenamefont
  {Rummukainen}, \citenamefont {Suorsa}, \citenamefont {Tuominen},\ and\
  \citenamefont {Tahtinen}}]{Leino:2017hgm}%
  \BibitemOpen
  \bibfield  {author} {\bibinfo {author} {\bibfnamefont {V.}~\bibnamefont
  {Leino}}, \bibinfo {author} {\bibfnamefont {K.}~\bibnamefont {Rummukainen}},
  \bibinfo {author} {\bibfnamefont {J.~M.}\ \bibnamefont {Suorsa}}, \bibinfo
  {author} {\bibfnamefont {K.}~\bibnamefont {Tuominen}}, \ and\ \bibinfo
  {author} {\bibfnamefont {S.}~\bibnamefont {Tahtinen}},\ }\href {\doibase
  10.1103/PhysRevD.97.114501} {\bibfield  {journal} {\bibinfo  {journal} {Phys.
  Rev.}\ }\textbf {\bibinfo {volume} {D97}},\ \bibinfo {pages} {114501}
  (\bibinfo {year} {2018}{\natexlab{a}})},\ \Eprint
  {http://arxiv.org/abs/1707.04722} {arXiv:1707.04722 [hep-lat]} \BibitemShut
  {NoStop}%
\bibitem [{\citenamefont {Leino}\ \emph
  {et~al.}(2018{\natexlab{b}})\citenamefont {Leino}, \citenamefont
  {Rummukainen},\ and\ \citenamefont {Tuominen}}]{Leino:2018qvq}%
  \BibitemOpen
  \bibfield  {author} {\bibinfo {author} {\bibfnamefont {V.}~\bibnamefont
  {Leino}}, \bibinfo {author} {\bibfnamefont {K.}~\bibnamefont {Rummukainen}},
  \ and\ \bibinfo {author} {\bibfnamefont {K.}~\bibnamefont {Tuominen}},\
  }\href@noop {} {\  (\bibinfo {year} {2018}{\natexlab{b}})},\ \Eprint
  {http://arxiv.org/abs/1804.02319} {arXiv:1804.02319 [hep-lat]} \BibitemShut
  {NoStop}%
\bibitem [{\citenamefont {Capitani}\ \emph {et~al.}(2006)\citenamefont
  {Capitani}, \citenamefont {Durr},\ and\ \citenamefont
  {Hoelbling}}]{Capitani:2006ni}%
  \BibitemOpen
  \bibfield  {author} {\bibinfo {author} {\bibfnamefont {S.}~\bibnamefont
  {Capitani}}, \bibinfo {author} {\bibfnamefont {S.}~\bibnamefont {Durr}}, \
  and\ \bibinfo {author} {\bibfnamefont {C.}~\bibnamefont {Hoelbling}},\ }\href
  {\doibase 10.1088/1126-6708/2006/11/028} {\bibfield  {journal} {\bibinfo
  {journal} {JHEP}\ }\textbf {\bibinfo {volume} {11}},\ \bibinfo {pages} {028}
  (\bibinfo {year} {2006})},\ \Eprint {http://arxiv.org/abs/hep-lat/0607006}
  {arXiv:hep-lat/0607006 [hep-lat]} \BibitemShut {NoStop}%
\bibitem [{\citenamefont {DeGrand}\ \emph {et~al.}(2011)\citenamefont
  {DeGrand}, \citenamefont {Shamir},\ and\ \citenamefont
  {Svetitsky}}]{DeGrand:2011vp}%
  \BibitemOpen
  \bibfield  {author} {\bibinfo {author} {\bibfnamefont {T.}~\bibnamefont
  {DeGrand}}, \bibinfo {author} {\bibfnamefont {Y.}~\bibnamefont {Shamir}}, \
  and\ \bibinfo {author} {\bibfnamefont {B.}~\bibnamefont {Svetitsky}},\
  }\bibfield  {booktitle} {\emph {\bibinfo {booktitle} {{Proceedings, 29th
  International Symposium on Lattice field theory (Lattice 2011): Squaw Valley,
  Lake Tahoe, USA, July 10-16, 2011}}},\ }\href@noop {} {\bibfield  {journal}
  {\bibinfo  {journal} {PoS}\ }\textbf {\bibinfo {volume} {LATTICE2011}},\
  \bibinfo {pages} {060} (\bibinfo {year} {2011})},\ \Eprint
  {http://arxiv.org/abs/1110.6845} {arXiv:1110.6845 [hep-lat]} \BibitemShut
  {NoStop}%
\bibitem [{\citenamefont {Rantaharju}\ \emph {et~al.}(2016)\citenamefont
  {Rantaharju}, \citenamefont {Rantalaiho}, \citenamefont {Rummukainen},\ and\
  \citenamefont {Tuominen}}]{Rantaharju:2015yva}%
  \BibitemOpen
  \bibfield  {author} {\bibinfo {author} {\bibfnamefont {J.}~\bibnamefont
  {Rantaharju}}, \bibinfo {author} {\bibfnamefont {T.}~\bibnamefont
  {Rantalaiho}}, \bibinfo {author} {\bibfnamefont {K.}~\bibnamefont
  {Rummukainen}}, \ and\ \bibinfo {author} {\bibfnamefont {K.}~\bibnamefont
  {Tuominen}},\ }\href {\doibase 10.1103/PhysRevD.93.094509} {\bibfield
  {journal} {\bibinfo  {journal} {Phys. Rev.}\ }\textbf {\bibinfo {volume}
  {D93}},\ \bibinfo {pages} {094509} (\bibinfo {year} {2016})},\ \Eprint
  {http://arxiv.org/abs/1510.03335} {arXiv:1510.03335 [hep-lat]} \BibitemShut
  {NoStop}%
\bibitem [{\citenamefont {Luscher}\ and\ \citenamefont
  {Weisz}(1996)}]{Luscher:1996vw}%
  \BibitemOpen
  \bibfield  {author} {\bibinfo {author} {\bibfnamefont {M.}~\bibnamefont
  {Luscher}}\ and\ \bibinfo {author} {\bibfnamefont {P.}~\bibnamefont
  {Weisz}},\ }\href {\doibase 10.1016/0550-3213(96)00448-8} {\bibfield
  {journal} {\bibinfo  {journal} {Nucl. Phys.}\ }\textbf {\bibinfo {volume}
  {B479}},\ \bibinfo {pages} {429} (\bibinfo {year} {1996})},\ \Eprint
  {http://arxiv.org/abs/hep-lat/9606016} {arXiv:hep-lat/9606016 [hep-lat]}
  \BibitemShut {NoStop}%
\bibitem [{\citenamefont {Luthe}\ \emph {et~al.}(2017)\citenamefont {Luthe},
  \citenamefont {Maier}, \citenamefont {Marquard},\ and\ \citenamefont
  {Schröder}}]{Luthe:2016xec}%
  \BibitemOpen
  \bibfield  {author} {\bibinfo {author} {\bibfnamefont {T.}~\bibnamefont
  {Luthe}}, \bibinfo {author} {\bibfnamefont {A.}~\bibnamefont {Maier}},
  \bibinfo {author} {\bibfnamefont {P.}~\bibnamefont {Marquard}}, \ and\
  \bibinfo {author} {\bibfnamefont {Y.}~\bibnamefont {Schröder}},\ }\href
  {\doibase 10.1007/JHEP01(2017)081} {\bibfield  {journal} {\bibinfo  {journal}
  {JHEP}\ }\textbf {\bibinfo {volume} {01}},\ \bibinfo {pages} {081} (\bibinfo
  {year} {2017})},\ \Eprint {http://arxiv.org/abs/1612.05512} {arXiv:1612.05512
  [hep-ph]} \BibitemShut {NoStop}%
\end{thebibliography}%
\bibliographystyle{apsrev4-1.bst}

\end{document}